\pdfoutput=1
\documentclass[a4paper,10pt]{llncs}

\usepackage[pdftex]{graphicx}
\usepackage{bsymb}
\usepackage[english]{babel}
\usepackage[utf8]{inputenc}
\usepackage{url}
\usepackage{array}
\usepackage{caption} 
\usepackage{multirow}
\usepackage{comment}
\usepackage{mathtools}
\usepackage[T1]{fontenc}
\usepackage{times}
\usepackage{changepage}	
\usepackage[final]{pdfpages}
\usepackage{tikz}
\usepackage{mathtools}
\usepackage{pbox}
\usepackage{amssymb}
\usepackage{graphicx}
\usepackage{float}
\usepackage{setspace}
\usepackage{listings}

\newcommand*\circled[1]{\tikz[baseline=(char.base)]{
            \node[shape=circle,color=white,fill=black,draw,inner sep=0.1pt] (char) {#1};}}
\newcommand{\minus}{\scalebox{0.75}[1.0]{$-$}}            
\newcommand{\negjoinrel}{\mathrel{\mkern3mu}}
\newcommand{\rlaprel}[1]{\mathrel{\mathrlap{#1}}}
\newcommand{\tsrel}{\rlaprel{\leftarrow}\negjoinrel\rlaprel{\leftarrow}
                    \joinrel\rlaprel{\rightarrow}\negjoinrel\rightarrow}

\makeatletter
\let\@fnsymbol\@alph
\makeatother

\title{From Event-B to Verified C via HLL}
\author{Ning Ge\inst{1}
\fnmsep \thanks{Seconded
from Systerel, Toulouse, France. \email{ning.ge@systerel.fr}}
\and
Arnaud Dieumegard \inst{1} 
\and Eric Jenn \inst{1}
\fnmsep \thanks{Seconded from Thales Avionics, Toulouse, France. \email{eric.jenn@fr.thalesgroup.com}}
\and Laurent Voisin \inst{2}
\institute{IRT-Saint Exupéry, Toulouse, France \\
\email{ning.ge|arnaud.dieumegard|eric.jenn@irt-saintexupery.fr}
\and
Systerel, Aix-en-provence, France \\
\email{laurent.voisin@systerel.fr}
}
}
\date{\today}

\begin{document}
\maketitle

\begin{abstract}
This work addresses the correct translation of an Event-B model to C code via an intermediate formal language, HLL. The proof of correctness follows two main steps. First, the final refinement of the Event-B model, including invariants, is translated to HLL. At that point, additional properties (e.g., deadlock-freeness, liveness properties, etc.) are added to the HLL model. The proof of the invariants and additional properties at the HLL level guarantees the correctness of the translation. Second, the C code is automatically generated from the HLL model for most of the system functions and manually for the remaining ones; in this case, the HLL model provides formal contracts to the software developer. An equivalence proof between the C code and the HLL model guarantees the correctness of the code.
\end{abstract}


\section{Introduction}
\label{sec:intro} 
Event-B \cite{abrial2010modeling} is a formal notation and method for the \textit{correct-by-construction} development of systems. In this method, a system is developed through a sequence of refinements, the consistency of which is formally proved. Event-B is based on first-order logic, typed set theory and integer arithmetic. As an integrated design environment for Event-B, the Rodin platform \cite{coleman2005rodin} provides support for refinement and mathematical proof of the consistency between refinements and of system-specific properties such as safety properties. In this work, we consider that the last refinement developed and proved with Event-B is the starting point for software development: the Event-B model is used as a specification for the implementation. 


Translating a correct specification to a correct implementation is challenging. Existing works \cite{wright2009automatic,bostrom2010creating,mery2011automatic,edmunds2011tasking,meryM13transformation,bostrom2014derivation,furst2014code} have identified the following main issues for this translation, and have proposed partial solutions. (1) It is necessary to restrict the Event-B model to a well-defined subset in order to generate code for a particular programming language. (2) When multiple events are enabled, then an event is chosen non-deterministically to be executed. A particular schedule needs to be defined to remove such non-determinism. (3) The correctness of translation needs to be guaranteed, which implies that the translation must preserve the safety properties expressed in the Event-B model. (4) Event-B does not come with constructs close to programming languages, as the B method \cite{abrial2005b} does with B0. Even though such constructs may be “emulated” by Event-B, the most refined event-B model usually only specifies contracts for software functions that need to be developed or provided outside Event-B (such as the implementation of the set theory). This leads to a verification gap between the Event-B contracts and the external or third-party implementation. (5) At the time of writing, it is still difficult to verify the deadlock-freeness and liveness properties using Rodin due to the difficulties of integrating loop variant reasoning in the Event-B model. These properties may be easier to verify using an intermediate verification language before the code is implemented. 
Among the above challenges, (1) and (2) have been addressed by all the existing works, while (3), (4) and (5) are still open. In this work, our focus is placed on the three open issues. A detailed discussion about the related works is provided in Section \ref{sec:related}.

Unlike other Event-B to code translation approaches, ours relies on an intermediate verification language, namely HLL (High Level Language). HLL is a synchronous declarative language similar to Lustre \cite{halbwachs1991synchronous}. It is the modeling language used by the formal verification toolset S3 (Systerel Smart Solver)\footnote{S3 is maintained, developed and distributed by Systerel (http://www.systerel.fr/).}\cite{clabaut2016industrial}, built around a SAT-based  model checker (S3-core). 
More details about the HLL language and the S3 toolset are presented in Section \ref{s3}. 

\begin{figure}[ht!]
	\centering
	\includegraphics[width=11cm]{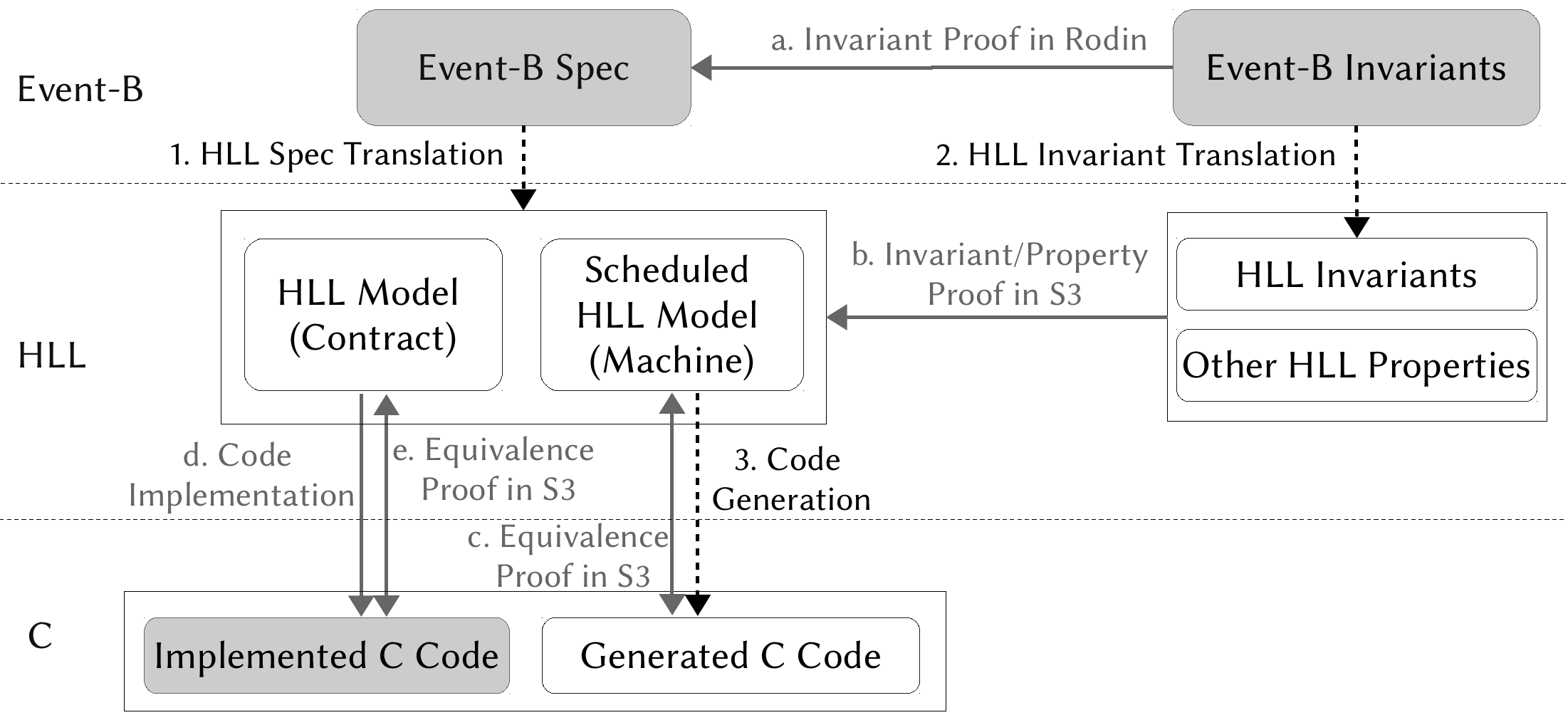}
       \caption{Process and Activities from Event-B to C via HLL}
       \label{fig:process}
\end{figure}

The overall translation process from Event-B to C via HLL is depicted in Figure \ref{fig:process}. On this figure, activities performed by existing tools are tagged by a letter while our contributions are tagged by a number. Activities are briefly explained hereafter.
\begin{itemize}
\item
	(a) When the Event-B refinement process is complete, all invariants in the model are proven in Rodin.
\item
	(1) The Event-B model is translated to HLL models that contain both a scheduled machine and a set of function con	tracts. Thanks to the function contracts, all invariants and additional properties are verifiable at the HLL level. This contribution will be detailed in Section \ref{sec:b2hll}.
\item
	(2) The invariants in Event-B are translated to HLL invariants. Some properties that are not verified in the Event-B model (e.g., deadlock-freeness and liveness ones) are expressed in HLL. This contribution will be detailed in Section \ref{sec:b2hll}.
\item
	(b) Compliance of the scheduled HLL model with the invariants and properties is proved using S3. If all invariants and properties are proven, the correctness of the HLL model is guaranteed. This activity will be detailed in Section \ref{sec:b2hll}.
\item
	(3) + (c) The C code of most system functions is generated from the scheduled HLL model. An equivalence proof is performed between the HLL model and the code to guarantee the correctness of the generated code. This contribution will be detailed in Section \ref{sec:hll2c}.
\item
	(d) + (e) Some functions of the C code are manually implemented from the function contracts in HLL. An equivalence proof is performed between the HLL contracts and the code to prove the correctness of the manually implemented code. This activity will be detailed in Section \ref{sec:hll2c}.
\end{itemize}

The organization of the paper is as follows: 
Section \ref{sec:related} discusses the related works; 
Section \ref{sec:background} describes the technological background, including the Event-B modeling language, the HLL modeling language and the S3 toolset, and presents our running example; 
Section \ref{sec:b2hll} illustrates the translation from the Event-B model to the HLL model;
Section \ref{sec:hll2c} exposes the translation from the HLL model to the C code; 
Section \ref{sec:arp} gives the experimental results on a significant use case: an automatic protection system for a robot;  
Section \ref{sec:conclusion} gives some concluding remarks and discusses perspectives.

\section{Related Works}
\label{sec:related}
As identified in Section \ref{sec:intro}, there are five main challenges for the translation from the Event-B model to the code. The issues of model restriction and prevention of non-determinism by scheduling have been solved in the existing works \cite{wright2009automatic,bostrom2010creating,mery2011automatic,edmunds2011tasking,meryM13transformation,bostrom2014derivation,furst2014code}. The correctness of the translation has been addressed in \cite{wright2009automatic,bostrom2010creating,mery2011automatic,bostrom2014derivation}. 
The author of \cite{mery2011automatic} intended to verify the generated and implemented code using meta-proof and software model checking tools such as BLAST \cite{beyer2007software}, which can check temporal safety properties of C program. As the approach was not yet experimented in their work and the details on the meta-proof and the checking of properties were not provided, it is difficult to evaluate the limits of this proposal.
The work \cite{meryM13transformation} generates verified C\# code in a static program verification environment, namely the Spec\# programming system \cite{barnett2004spec}, that is based on deductive verification of function contracts. 
For C code, the same idea can be applied using the Frama-C framework \cite{cuoq2012frama}. However, the verification based on function contracts is restricted to local properties. It is not easy to express and verify the safety properties concerning the global behavior of a system, as well as the deadlock-freeness and liveness properties. 
The work \cite{furst2014code} generates correct C code relying on reasoning about well-definedness, assertions and refinement. It relies on a set of well-definedness restriction rules for the Event-B model to prevent the occurrence of runtime errors such as arithmetic overflows. The correctness of the generated C code is informally guaranteed by using a refinement step. To verify that the control flow is the same as the restricted Event-B model, this work adds new variables that represents the program counter in the scheduled model and adds events that simulate the update of the counter. This approach is restricted to the verification of control flow invariants. In addition, it does not yet cover deadlock-freeness and liveness properties. 

\section{Technological Background}
\label{sec:background}

\subsection{The Event-B Modeling Language and the Rodin Platform}
\label{eventb}	
Event-B is a modeling formalism and method for the development of systems relying on a correct-by-construction approach. At the time of writing, Event-B focuses on discrete transition systems. Centered around the notion of \textit{events}, it is structured around \textit{contexts} and \textit{machines}. A \textit{context} defines the type of data and models the static properties of the system including the function contracts. A \textit{machine} models the dynamic behavior by means of variables, the values of which are initially determined by the \textit{initialization} event and changed by actions of \textit{events}. 
Construction \textit{invariants} and safety properties are expressed in the machines. 
An \textit{event} waits for a set of guard conditions to be verified to trigger a sequence of \textit{actions}. A \textit{parameter} is constrained by appropriate \textit{guards} and its chosen value is used to update the machine \textit{variables}. A \textit{witness} maps a concrete parameter with a more abstract one defined in a less refined machine. Details about the Event-B language and method can be found in \cite{abrial2010modeling}.

\subsection{The HLL Modeling Language and the S3 Toolset}	
\label{s3}
\begin{figure}[ht!]
	\centering
	\includegraphics[width=12.5cm]{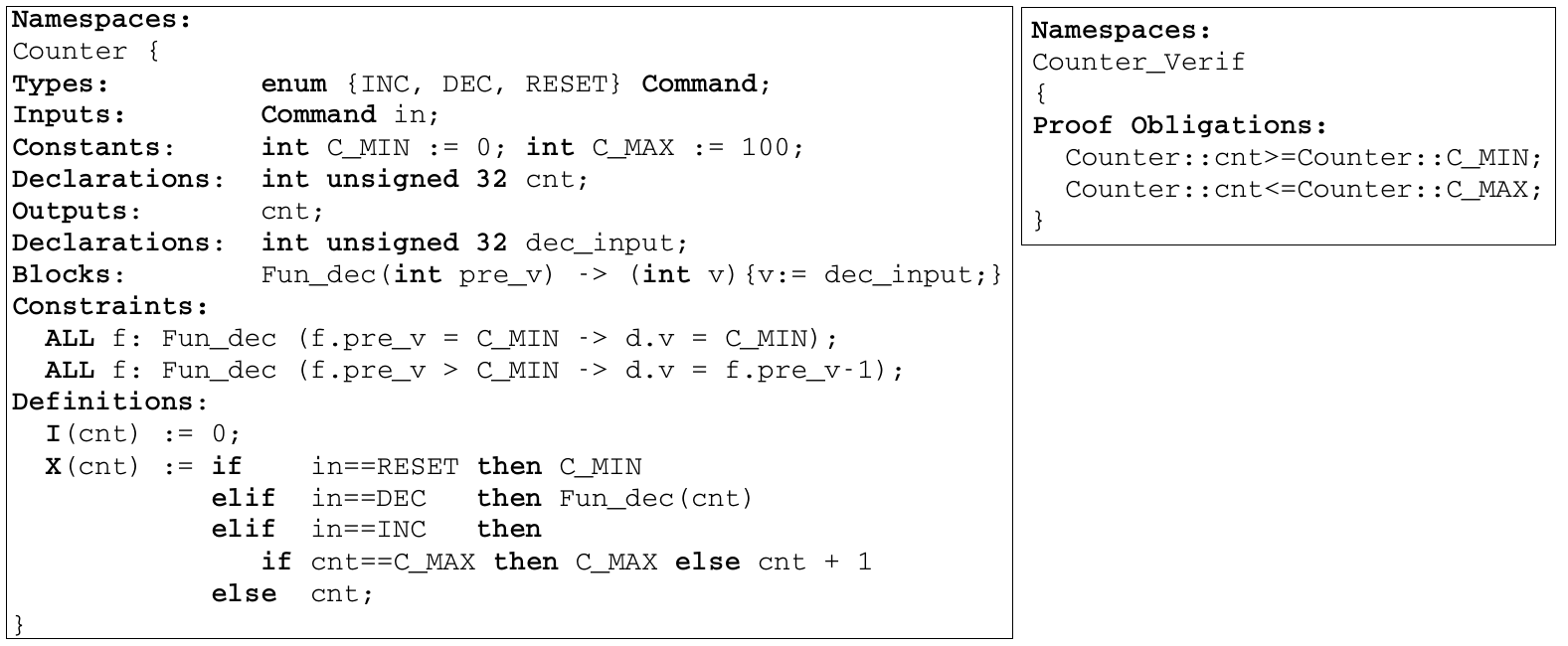}
\caption{An Example of HLL Model}
\label{fig:hll}
\end{figure}

HLL is a synchronous dataflow language used to model a system, its environmental constraints as well as its properties. To give an overview of the language constructs, Figure \ref{fig:hll} shows the HLL model of a saturated counter and its property about the range of output value (respectively in the \textit{namespaces} $\mathtt{Counter}$ and $\mathtt{Counter\_Verif}$).
The counter reacts to the input command (modelled as an HLL enumeration): incrementation ($\mathtt{INC}$), decrementation ($\mathtt{DEC}$) or reset ($\mathtt{RESET}$). The saturation range is defined by HLL \textit{constants}. The behavior of the counter is initialized by $\mathtt{\textbf{I}(cnt)}$ and periodically updated by $\mathtt{\textbf{X}(cnt)}$. The effect of $\mathtt{INC}$ and $\mathtt{RESET}$ are directly defined in the schedule, while the effect of $\mathtt{DEC}$ is defined as a function contract by using HLL \textit{constraints} and an intermediate variable $\mathtt{dec\_input}$ of the HLL \textit{block} $\mathtt{Fun\_dec()}$, without any concrete implementation.

HLL is used as the modeling/verification language of the S3 toolset. Design models specified in SCADE \cite{caspi2003simulink}/Lustre \cite{halbwachs1991synchronous} or code in C/Ada can be translated to HLL thanks to translators provided by the toolset. The proof engine of S3 is a SAT-based \cite{biere2009handbook} model checker, S3-core, that implements Bounded Model Checking (BMC) \cite{biere1999symbolic} and k-induction \cite{sheeran2000checking,bjesse2000sat} techniques. The input of S3-core is a bit-level Low Level Language (LLL) that only contains boolean streams and is restricted to three bitwise operators: negation, implication and equivalence. The toolset provides "expanders" to translate HLL models into LLL models. More details about HLL and S3 can be found in \cite{ge2016formal}. 

S3 supports different activities of a software development process: property proof, equivalence proof, automatic test case generation, simulation, and provides necessary elements to comply with the software certification processes. It has been used for the formal verification of railway signalling systems for years by various industrial companies in this field. 

\subsection{A Running Example: A Traffic Light Controller}
\label{example}
To make this paper more readable and understandable, we use the case of the traffic light controller proposed in the Rodin User’s Handbook\footnote{https://www3.hhu.de/stups/handbook/rodin/current/html/} as a running example. In order to illustrate the role of function contracts, we have slightly modified the original Event-B model by abstracting an event action into a set of function interface and the contracts of the function. The requirements of this traffic light controller are the follows:
\begin{itemize}
\item
	\texttt{REQ-1}: A traffic light controller shall be used to control both the pedestrian traffic lights and car traffic lights of a pedestrian crossing.
\item
	\texttt{REQ-2}: Both traffic lights shall not be \textit{green} at the same time, to ensure that the pedestrian crossing is in a safe state.
\end{itemize}

This example covers most Event-B constructs. The model is provided in the Appendix A. It is composed of an abstract machine ($\mathtt{mac0}$) without context and a refined machine ($\mathtt{mac1}$) with a context($\mathtt{ctx1}$). The specified invariants are related to the data type, the usage of gluing variables, and the safety property (\texttt{REQ-2}). The contract of the  function that sets the pedestrian light to red color is specified in the context $\mathtt{ctx1}$.

\section{Translating Event-B to Verified HLL}
\label{sec:b2hll}

\subsection{Architecture of the HLL Model}
\label{sec:arch_hll}
\begin{figure}[ht!]
	\begin{center}
	\includegraphics[width=11cm]{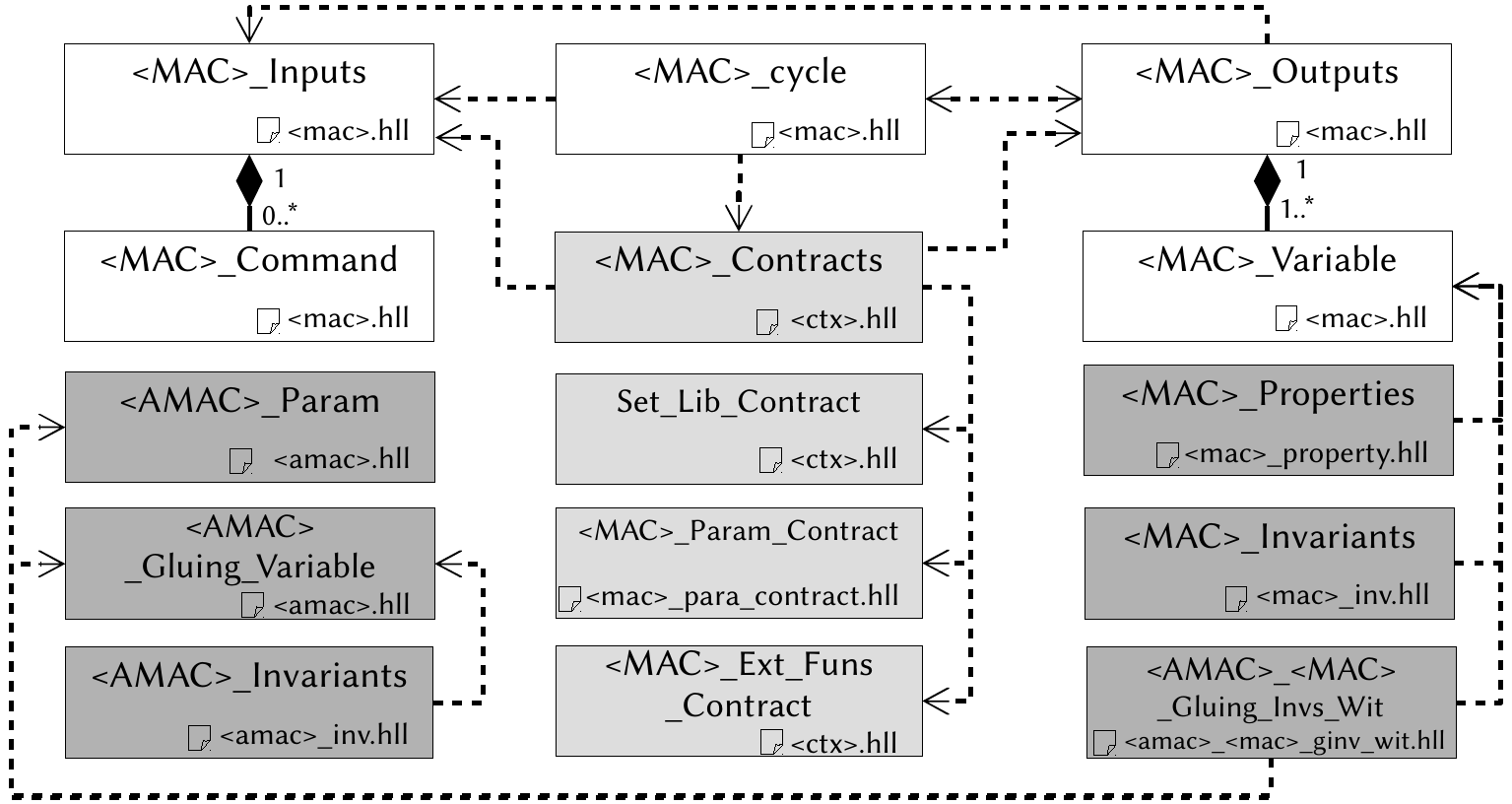}
    \caption{Static Architecture of the HLL Model}
    \label{fig:arch_hll}
    \end{center}
\end{figure}

Figure \ref{fig:arch_hll} depicts the static architecture of the HLL model. A complete model is composed of three main parts: (i) the specification of the scheduled system (the white blocks), (ii) the specification of the invariants/properties and other verification related artefacts (the dark grey blocks), and (iii) the function contracts (the light grey blocks). Code generation requires part (i); the implementation of contracted functions requires part (iii); and the verification of the HLL model requires all parts. The HLL model of the system consists of the \textit{inputs} of system \textit{commands}, the \textit{outputs} of the Event-B \textit{variables} of the machine, the function \textit{contracts} for the parameters, external functions and third-party library (e.g., the implementation of the set theory and HLL quantifiers). The verification architecture consists of the invariants/properties, the gluing invariants/witnesses in the less refined machine, and the related gluing variables/parameters in the abstract events. 

Concerning the dynamic architecture of the model, the main question concerns the scheduling of events. When multiple events are enabled, then an event is chosen non-deterministically to be executed. However, to implement such model with a sequential programming language, and to ensure the deterministic behavior of the software, a specific scheduling of events must be chosen. The problem of event scheduling has been addressed in many existing works. We are not aimed to enumerate the translation of every possible scheduling algorithms. As an example, we use a simple algorithm where events are processed in their definition order in the Event-B model. 

\subsection{Translation Rules from Event-B to HLL}
We present the translation rules according to the presentation of the Event-B constructs in \cite{robinson2009concise}, including arithmetic operations, predicates, sets, and relations.

\subsubsection{Translating Arithmetic and Predicates}
The translation rules of arithmetic and predicates are defined in Table \ref{table:B_arth_pred}. These translations are straightforward, thanks to the quantifiers in HLL language. Note that the quantifiers are executable at the HLL level, but they need to be implemented at the code level.
\begin{table}[ht!]
\caption{Event-B Arithmetic and Predicates to HLL Model}
\footnotesize
\begin{tabular}{c c}
\begin{tabular}[t]{ l | l | l }
  \textbf{Arithmetic} & \textbf{Event-B} & \textbf{HLL} \\
  \hline
  Integers & $\mathtt{INT}$ &  $\mathtt{int \enspace signed \enspace N }$\\
  Natural Number & $\mathtt{NAT}$ & $\mathtt{int \enspace unsigned \enspace N }$ \\
  Interval & $\mathtt{x = n..m}$ & $\mathtt{int[n,m] \enspace x;}$ \\
  Addition & $\mathtt{x := a + b}$ & $\mathtt{x := a + b;}$ \\
  Subtraction & $\mathtt{x := a - b}$ & $\mathtt{x := a - b;}$ \\
  Multiplication & $\mathtt{x := a * b}$ & $\mathtt{x := a * b;}$ \\
  Division & $\mathtt{x := a \div b}$ & $\mathtt{x := a \enspace / \enspace b;}$ \\
  Modulo & $\mathtt{x := a \enspace mod \enspace b}$ & $\mathtt{x := a \enspace \% \enspace b;}$ \\
  Exponentiation & $\mathtt{x := a{\mathchar"5E}b}$ & $\mathtt{x := a\mathchar"5E b;}$ \\
  Minimum & $\mathtt{min(S)}$ & $\mathtt{\$min \, i:[0,N \minus 1] (S(i));}$ \\
  Maximum & $\mathtt{max(S)}$ & $\mathtt{\$max \, i:[0,N \minus 1] (S(i));}$   \\
\end{tabular} &
\begin{tabular}[t]{ l | l | l }
  \textbf{Predicate} & \textbf{Event-B} & \textbf{HLL} \\
  \hline
  Negation & $\mathtt{\neg x}$ & $\mathtt{{\raise.17ex\hbox{$\scriptstyle\sim$}} x}$ \\
  True  & $\mathtt{\top}$ & $\mathtt{true}$ \\
  False & $\mathtt{\bot}$ & $\mathtt{false}$ \\  
  Equality & $\mathtt{a = b}$ & $\mathtt{a == b}$ \\ 
  Inequality & $\mathtt{a \neq b}$ & $\mathtt{a \enspace != b}$ \\
  Less &$\mathtt{a < b}$ & $\mathtt{a < b}$ \\
  Less or equal &$\mathtt{a \leq b}$ & $\mathtt{a \leq b}$ \\
  Greater &$\mathtt{a > b}$ & $\mathtt{a > b}$ \\
  Greater or equal &$\mathtt{a \geq b}$ & $\mathtt{a \geq b}$ \\    
  Conjunction & $\mathtt{x \wedge y}$ & $\mathtt{x \enspace \scriptstyle\&  \textstyle \enspace y}$\\
  Disjunction & $\mathtt{x \vee y}$ & $\mathtt{x \enspace \scriptstyle\# \textstyle \enspace y}$ \\
  Implication & $\mathtt{x \Rightarrow y}$ & $\mathtt{x \rightarrow y}$ \\
  Equivalence & $\mathtt{x \Leftrightarrow y}$ & $\mathtt{x \leftrightarrow y}$ \\
  \hline
  \pbox[c]{30cm}{Universal \\ quantification } & $\forall$ & $\mathtt{ALL}$ \\
  \hline
  \pbox[c]{20cm}{Existential \\ quantification} & $\exists$ & $\mathtt{SOME}$ \\
\end{tabular}
\tabularnewline
\end{tabular}
\label{table:B_arth_pred}
\end{table}
	
\subsubsection{Translating Set Theory}	
\label{sec:hll_set}
The translation rules of sets and set predicates are defined in Table \ref{table:B_sets} and Table \ref{table:B_set_pred}. They are implemented using the characteristic functions in HLL.
In the HLL model, a set $\mathtt{S}$ of finite size $\mathtt{N}$ with elements $\mathtt{e_i (i = 0 .. N \minus 1)}$ is implemented as a boolean array $\mathtt{A}$ of size $\mathtt{N}$ such that $\mathtt{S[i]}$ is true if and only if element $\mathtt{e_i}$ is in $\mathtt{S}$.  This implementation has strong restrictions. In particular, it imposes that the cardinality of $\mathtt{S}$ is bounded (and “reasonnable”). Many other implementations would have been possible, nevertheless, we chose this one to leverage the optimized operations on boolean vectors implemented in S3. Note that the code implementation of the HLL set constraints need to be provided by a set theory library.  

\begin{table}[ht!]
\caption{Event-B Sets to HLL Model}
\centering
\footnotesize
\begin{tabular}{ l | l | l }
  \textbf{Sets} & \textbf{Event-B} & \textbf{HLL} \\
  \hline
  Set & $\mathtt{S := \{e_1, ... , e_N\}}$ & $\mathtt{bool \enspace S[N];}$ \\
  Empty set & $\mathtt{S := \emptyset}$ & $\mathtt{S[i] := false;}$ \\
  Set comprehension & $\mathtt{E := \{z \cdot P \mid F \}}$ & $\mathtt{E[z] :=  F[z] \enspace \scriptstyle\& \textstyle \enspace P(z);}$\\
  Union & $\mathtt{U := S \bunion T}$ & $\mathtt{U[i] := S[i] \enspace \scriptstyle\# \textstyle \enspace T[i];}$ \\
  Intersection & $\mathtt{U := S \binter T}$ & $\mathtt{U[i] := S[i] \enspace \scriptstyle\& \textstyle \enspace T[i];}$ \\
  Difference & $\mathtt{U := S \setminus T}$ & $\mathtt{U[i] := S[i] \enspace \scriptstyle\& \textstyle \enspace {\raise.17ex\hbox{$\scriptstyle\sim$}} T[i];}$ \\
  Cartesian product & $\mathtt{E := S \cprod T}$ & $\mathtt{E[i][j] := S[i] \enspace \scriptstyle\& \textstyle \enspace T[j];}$ \\ 
  \hline
  Powerset & $\pow\mathtt{(S)}$ & \pbox{20cm}{$\mathtt{Declarations: bool \,\, P[N];}$  \\ 
 $\mathtt{Constraints:}$ \\
 $\mathtt{ \enspace ALL \enspace i:[0, N\minus 1] \enspace ({\raise.17ex\hbox{$\scriptstyle\sim$}}S[i] \rightarrow {\raise.17ex\hbox{$\scriptstyle\sim$}}P[i] );}$ 
  }\\
  \hline
  Non-empty subsets & $\pown \mathtt{(S)}$ & \pbox{20cm}{$\mathtt{Declarations:\enspace bool \enspace P[N];}$ \\ 
$\mathtt{Constraints:}$  \\
 $\mathtt{ \enspace ALL \enspace i:[0, N\minus 1] \enspace ({\raise.17ex\hbox{$\scriptstyle\sim$}}S[i] \rightarrow {\raise.17ex\hbox{$\scriptstyle\sim$}}P[i] );}$ \\
$\mathtt{ \enspace SOME \enspace i:[0,N \minus 1] \enspace (S[i] \rightarrow  {\raise.17ex\hbox{$\scriptstyle\sim$}}P[i]) ;}$ }\\
  \hline
  Cardinality & $\mathtt{card(S)}$ & $\mathtt{card := population(S);}$ \\
  Generalized union & $\mathtt{union(S)}$ & $\mathtt{U[i] := SOME \enspace j:[0,N \minus 1] \enspace (S[i][j]);}$ \\
  Generalized intersection & $\mathtt{inter(S)}$ & $\mathtt{U[i] := ALL \enspace j:[0,N \minus 1] \enspace (S[i][j]);}$ \\  
  Quantified union & $\mathtt{UNION \enspace z\cdot P | S}$ & $\mathtt{U[i] := SOME \enspace j:[0,N \minus 1] \enspace (P(S[i][j]));}$ \\
  Quantified intersection & $\mathtt{INTER\, z\cdot P | S}$ & $\mathtt{U[i] := ALL \enspace j:[0,N \minus 1] \enspace (P(S[i][j]));}$ \\  
\end{tabular}
\label{table:B_sets}
\end{table}	

\begin{table}[ht!]
\caption{Event-B Set Predicates to HLL Model}
\centering
\footnotesize
\begin{tabular}{ l | l | l }
  \textbf{Set Predicates} & \textbf{Event-B} & \textbf{HLL} \\
  \hline
  Set membership & $\mathtt{s \enspace : \enspace S}$ & $\mathtt{S(s) == true;}$ \\
  Set non-membership & $\mathtt{s \enspace /: \enspace S}$ & $\mathtt{S(s) == false;}$ \\
  \hline
  Subset & $\mathtt{S \enspace \subseteq \enspace T}$ & $\mathtt{ ALL \enspace i:[0,N\minus 1] \enspace (S[i] \rightarrow T[i] );}$\\
  \hline
  Proper subset & $\mathtt{S \enspace \subset \enspace T}$ & \pbox{20cm}{$\mathtt{ALL \enspace i:[0,N\minus 1] \enspace (S[i] \rightarrow T[i]) \enspace \scriptstyle\&\textstyle }$ \\
$\mathtt{ SOME \enspace i:[0,N \minus 1] \enspace (T[i] \rightarrow  {\raise.17ex\hbox{$\scriptstyle\sim$}}S[i]) ;}$  } \\ 
  \hline
  Finite set & $\mathtt{finite(S)}$ & (All sets in HLL are finite.) \\
  Partition & $\mathtt{partition(S, s_1,...,s_n)}$ & $\mathtt{S[i] := s_1[i] \enspace \scriptstyle\#\textstyle \enspace ... \enspace \scriptstyle\#\textstyle \enspace s_n[i];}$ \\
\end{tabular}
\label{table:B_set_pred}
\end{table}	

\subsubsection{Translating Relations and Functions}
A relation is a set of ordered pairs. The translation rules of relations are defined in Table \ref{table:B_relations}. As a set in Event-B is implemented as a boolean array in HLL, a relation $\mathtt{r}$ of two sets $\mathtt{S}$ of size $\mathtt{N}$ and $\mathtt{T}$ of size $\mathtt{M}$ is implemented as a 2-dimensional boolean array $\mathtt{R[N][M]}$. The HLL pre-conditions of a relation $\mathtt{r}$ between elements $\mathtt{s_i}$ of $\mathtt{S}$ and $\mathtt{t_j}$ of $\mathtt{T}$ is (i) $\mathtt{s_i}$ are in $\mathtt{S}$ ($\mathtt{S[i]}$ is true), (ii) $\mathtt{t_j}$ is in $\mathtt{T}$ ($\mathtt{T[j]}$ is true), (iii) $\mathtt{r}$ exists between $\mathtt{s_i}$ and $\mathtt{t_j}$ ($\mathtt{R[i][j]}$ is true). 

In Event-B, a special case of relations are functions, with the restriction that each element of the domain is related to a unique element in the range. The translation rules of functions are defined in Table \ref{table:B_functions}. 

\begin{table}[ht!]
\caption{Event-B Relations to HLL Model}
\centering
\footnotesize
\begin{tabular}{ l | l | l }
  \textbf{Relations} & \textbf{Event-B} & \textbf{HLL} \\
  \hline
  Relations & $\mathtt{S \rel T}$ & \pbox{20cm}{$\mathtt{Declarations: bool \,\, r[N][M];}$ \\ $\mathtt{Constraints :}$ \\ $\mathtt{ \enspace ALL \enspace i:[0,N \minus 1], j:[0,M \minus 1] \Big( ({\raise.17ex\hbox{$\scriptstyle\sim$}}S[i] \scriptstyle\# \textstyle {\raise.17ex\hbox{$\scriptstyle\sim$}}T[j]) \rightarrow {\raise.17ex\hbox{$\scriptstyle\sim$}}r[i][j] \Big);}$} \\
  \hline
  Domain & $\mathtt{dom(r)}$ & $\mathtt{dom[i] := SOME \enspace j:[0,N \minus 1] \enspace (r[i][j]);}$  \\
  Range & $\mathtt{ran(r)}$ &  $\mathtt{ran[j] := SOME \enspace i:[0,N \minus 1] \enspace (r[i][j]);}$  \\
  \hline
  Total relation & $\mathtt{S \trel T} $ & \pbox{20cm}{$\mathtt{Declarations: bool \enspace r[N][M];}$ \\ 
  $\mathtt{Constraints :}$ \\ 
  $\mathtt{ \enspace ALL \enspace i:[0,N \minus 1] \enspace \big( S[i] \rightarrow SOME \enspace j:[0,M \minus 1] \enspace (T[j] \enspace \scriptstyle\& \textstyle \enspace r[i][j]) \big) ;}$ \\ 
  $\mathtt{ \enspace ALL \enspace i:[0,N \minus 1] \enspace \big( {\raise.17ex\hbox{$\scriptstyle\sim$}}S[i] \rightarrow ALL \enspace j:[0,M \minus 1] \enspace ({\raise.17ex\hbox{$\scriptstyle\sim$}}r[i][j]) \big);}$} \\
  \hline
  \pbox{20cm}{Surjective \\Relation }& $\mathtt{S \srel T}$ & \pbox{20cm}{$\mathtt{Declarations: bool \enspace r[N][M];}$ \\ $\mathtt{Constraints :}$ \\ 
  $\mathtt{ \enspace ALL \enspace j:[0,M \minus 1] \enspace \big( T[i] \rightarrow SOME \enspace i:[0,N \minus 1] \enspace (S[i] \enspace \scriptstyle\& \textstyle \enspace r[i][j]) \big) ;}$ \\ 
  $\mathtt{ \enspace ALL \enspace i:[0,M \minus 1] \enspace \big( {\raise.17ex\hbox{$\scriptstyle\sim$}}T[j] \rightarrow ALL \enspace i:[0,N \minus 1] \enspace ({\raise.17ex\hbox{$\scriptstyle\sim$}}r[i][j]) \big);}$} \\
  \hline
  \pbox{20cm}{Total surjective \\ relation} & $\mathtt{S \tsrel T}$ & \pbox{20cm}{$\mathtt{Declarations: bool \enspace r[N][M];}$ \\ $\mathtt{Constraints :}$ \\ 
$\mathtt{ \enspace ALL \enspace i:[0,N \minus 1] \enspace \big( S[i] \rightarrow SOME \enspace j:[0,M \minus 1] \enspace (T[j] \enspace \scriptstyle\& \textstyle \enspace r[i][j]) \big) ;}$ \\ 
  $\mathtt{ \enspace ALL \enspace j:[0,M \minus 1] \enspace \big( T[i] \rightarrow SOME \enspace i:[0,N \minus 1] \enspace (S[i] \enspace \scriptstyle\& \textstyle \enspace r[i][j]) \big) ;}$ \\ 
  $\mathtt{ \enspace ALL \enspace i:[0,N \minus 1] \enspace \big( {\raise.17ex\hbox{$\scriptstyle\sim$}}S[i] \rightarrow ALL \enspace j:[0,M \minus 1] \enspace ({\raise.17ex\hbox{$\scriptstyle\sim$}}r[i][j]) \big);}$ \\
  $\mathtt{ \enspace ALL \enspace i:[0,M \minus 1] \enspace \big( {\raise.17ex\hbox{$\scriptstyle\sim$}}T[j] \rightarrow ALL \enspace i:[0,N \minus 1] \enspace ({\raise.17ex\hbox{$\scriptstyle\sim$}}r[i][j]) \big);}$} \\

\end{tabular}
\label{table:B_relations}
\end{table}	

\begin{table}[ht!]
\caption{Event-B Functions to HLL Model}
\centering
\footnotesize
\begin{tabular}{ l | l | l }
\textbf{Functions} & \textbf{Event-B} & \textbf{HLL} \\
\hline
Partial function & $\mathtt{f \in S \pfun T}$ & $\mathtt{ALL \enspace s: S, t_1, t_2: T \enspace \big(t_1 = f(s) \enspace \scriptstyle\&\textstyle \enspace t_2 = f(s) \rightarrow t_1 = t_2 \big);}$ \\ 
\hline
Total function & $\mathtt{f \in S \tfun T }$ & 
\pbox{20cm}{$\mathtt{ALL \enspace s: S, t_1, t_2: T \enspace \big(t_1 = f(s) \enspace \scriptstyle\&\textstyle \enspace t_2 = f(s) \rightarrow t_1 = t_2 \big);}$ \\
$\mathtt{ALL \enspace s:S \enspace \big(SOME \enspace t:T \enspace (f(s) = t) \big);}$
}\\
\hline
Partial injection & $\mathtt{f \in S \pinj T}$ & 
\pbox{20cm}{$\mathtt{ALL \enspace s: S, \enspace t_1, t_2: T \enspace \big(t_1 = f(s) \enspace \scriptstyle\&\textstyle \enspace t_2 = f(s) \rightarrow t_1 = t_2 \big);}$ \\
$\mathtt{ALL \enspace s_1, s_2:S, \enspace t: S \enspace \big( t = f(s_1) \enspace \scriptstyle\&\textstyle \enspace t = f(s_2) \enspace \rightarrow s_1 = s_2 \big);}$
} \\ 
\hline
Total injection & $\mathtt{f \in S \tinj T}$ & 
\pbox{20cm}{$\mathtt{ALL \enspace s: S, \enspace t_1, t_2: T \enspace \big(t_1 = f(s) \enspace \scriptstyle\&\textstyle \enspace t_2 = f(s) \rightarrow t_1 = t_2 \big);}$ \\
$\mathtt{ALL \enspace s_1, s_2:S, \enspace t: S \enspace \big( t = f(s_1) \enspace \scriptstyle\&\textstyle \enspace t = f(s_2) \enspace \rightarrow s_1 = s_2 \big);}$ \\
$\mathtt{ALL \enspace s:S \enspace \big(SOME \enspace t:T \enspace (f(s) = t) \big);}$
}\\
\hline
Partial surjection & $\mathtt{f \in S \psur T}$ & 
\pbox{20cm}{$\mathtt{ALL \enspace s: S, t_1, t_2: T \enspace \big(t_1 = f(s) \enspace \scriptstyle\&\textstyle \enspace t_2 = f(s) \rightarrow t_1 = t_2 \big);}$ \\
$\mathtt{ALL \enspace t:T \enspace \big(SOME \enspace s:S \enspace (f(s) = t) \big);}$
}\\
\hline
Total surjection & $\mathtt{f \in S \tsur T}$ & 
\pbox{20cm}{$\mathtt{ALL \enspace s: S, \enspace t_1, t_2: T \enspace \big(t_1 = f(s) \enspace \scriptstyle\&\textstyle \enspace t_2 = f(s) \rightarrow t_1 = t_2 \big);}$ \\
$\mathtt{ALL \enspace s:S \enspace \big(SOME \enspace t:T \enspace (f(s) = t) \big);}$ \\
$\mathtt{ALL \enspace t:T \enspace \big(SOME \enspace s:S \enspace (f(s) = t) \big);}$
}\\
\hline
Bijection & $\mathtt{f \in S \tbij T}$ & 
\pbox{20cm}{$\mathtt{ALL \enspace s: S, \enspace t_1, t_2: T \enspace \big(t_1 = f(s) \enspace \scriptstyle\&\textstyle \enspace t_2 = f(s) \rightarrow t_1 = t_2 \big);}$ \\
$\mathtt{ALL \enspace s_1, s_2:S, \enspace t: S \enspace \big( t = f(s_1) \enspace \scriptstyle\&\textstyle \enspace t = f(s_2) \enspace \rightarrow s_1 = s_2 \big);}$ \\
$\mathtt{ALL \enspace s:S \enspace \big(SOME \enspace t:T \enspace (f(s) = t) \big);}$ \\
$\mathtt{ALL \enspace t:T \enspace \big(SOME \enspace s:S \enspace (f(s) = t) \big);}$
}\\
\end{tabular}
\label{table:B_functions}
\end{table}

\subsubsection{Translating Event-B Constructs}
Table \ref{table:b_hll} presents the mapping between Event-B constructs and the HLL elements. 

\begin{table}[ht!]
\caption{Event-B Constructs to HLL Model}
\centering
\scriptsize
\begin{tabular}{ l | l | l }
  \textbf{Event-B Construct} & \textbf{HLL Construct} & \textbf{Dependence} \\
  \hline
  $\mathbf{CONTEXT}$ & $\mathbf{ctx.hll}$ & \\
  $\mathtt{EXTENDS}$ & $\mathtt{Abstract \enspace context \enspace HLL}$ &  $\mathtt{abs\_ctx.hll}$ \\
  $\mathtt{SETS}$ & $\mathtt{Enum, Array}$ & \\ 
  $\mathtt{CONSTANTS}$ & $\mathtt{Type}$ & \\
  $\mathtt{AXIOMS}$ & $\mathtt{Block, Constraint}$ & \\
  \hline
  $\mathbf{MACHINE}$ & $\mathbf{mac.hll}$ & \\
  $\mathtt{REFINES}$ & $\mathtt{Abstract \enspace machine \enspace HLL}$ & $\mathtt{abs\_mac.hll}$ \\
  $\mathtt{SEES}$ & $\mathtt{Abstract \enspace context \enspace HLL}$ & $\mathtt{abs\_ctx.hll}$ \\
  $\mathtt{VARIABLES}$ & $\mathtt{Inputs, Struct}$ &  \\
  $\mathtt{INVARIANT}$ & $\mathtt{Types, Proof \enspace obligations}$ & $\mathtt{mac.hll, (abs\_)mac\_inv.hll}$ \\
  $\mathtt{Gluing \enspace INVARIANT}$ & $\mathtt{Constraints}$ & $\mathtt{mac\_absmac\_ginv.hll}$ \\
  \hline
  $\mathbf{EVENT}$ & $\mathbf{mac.hll}$ & \\
  $\mathtt{REFINES}$ & $\mathtt{Abstract \enspace machine \enspace HLL}$ & $\mathtt{abx\_mac.hll}$ \\
  $\mathtt{ANY}$ & $\mathtt{Declarations}$ &  $\mathtt{mac\_para\_contract.hll}$ \\
  $\mathtt{WHERE}$ & $\mathtt{Predicate(vars)}$ & $\mathtt{mac\_para\_contract.hll}$ \\
  $\mathtt{WITH}$ & $\mathtt{Constraints}$ & $\mathtt{mac\_absmac\_ginv.hll}$ \\
  $\mathtt{THEN}$ & $\mathtt{Definitions, I(vars), X(vars)}$ &   \\
\end{tabular}
\label{table:b_hll}
\end{table}

\begin{figure}[ht!]
	\centering
	\includegraphics[width=12.5cm]{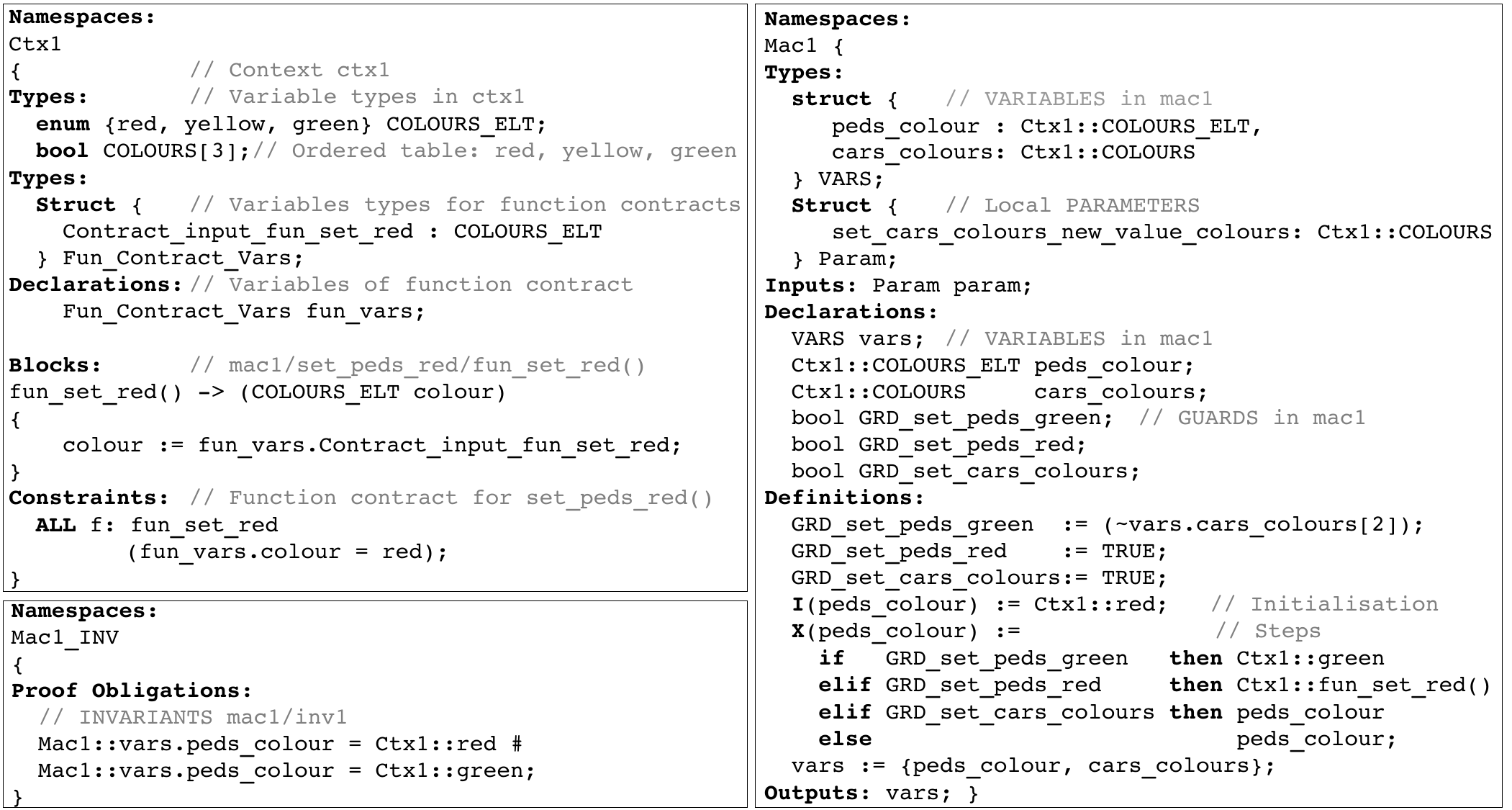}
    \caption{Partial HLL Models of the Running Example}
    \label{fig:mac1_hll}
\end{figure}

Part of HLL model of the running example is provided in Figure \ref{fig:mac1_hll}, including one context ($\mathtt{ctx1}$), one machine ($\mathtt{mac1}$), and the invariants in the machine $\mathtt{mac1}$ ($\mathtt{mac1\_INV}$). 
In the HLL model of machine $\mathtt{mac1}$, for the reason of space limitations, only the variable $\mathtt{peds\_colour}$ is defined to explain the schedule. The complete HLL model of the running example is provided in the Appendix B. As explained in Section \ref{sec:arch_hll}, a sequential schedule is applied in the HLL model. Firstly, the guards (i.e. $\mathtt{GRD\_set\_peds\_green}$, $\mathtt{GRD\_set\_peds\_red}$ and $\mathtt{GRD\_set\_cars\_colours}$) are defined. Then each variable is initialized by the definition $\mathtt{I(vars)}$, such as $\mathtt{I(peds\_colour)}$. The update of values with respect to the schedule of events is defined in the step $\mathtt{X(var)}$, such as $\mathtt{X(peds\_colour)}$.

\subsection{Proving HLL Invariants and Properties in S3}
The invariants in the Event-B models are translated to \textit{HLL Proof Obligations} (see example \texttt{Mac1\_INV} in Figure \ref{fig:mac1_hll}). The translation rules are the same as that defined in the previous sections. Besides the invariants of refinement, usually the other invariants in Event-B represent safety properties. The deadlock-freeness and liveness properties will be directly expressed in HLL. The liveness property is handled by using lasso \cite{biere2002liveness,hoang2011reasoning}. The deadlock-freeness \cite{yang2011event} property is expressed using the guards of events. The HLL expression of the deadlock-freeness property is provided here. Note that $\mathtt{a := true, b;}$  is another written form for $\mathtt{I(a) := true; \enspace X(a) := b;}$. 
\begin{align*}
\mathtt{PROP\_DLF := \enspace true, \enspace mac::GRD\_E_1 \enspace \scriptstyle\#\textstyle \enspace mac::GRD\_E_2 \enspace \scriptstyle\#\textstyle \enspace ...\enspace  \scriptstyle\#\textstyle \enspace mac::GRD\_E_n;}
\end{align*}

\begin{figure}[ht!]
		\begin{center}
		\includegraphics[width=9cm]{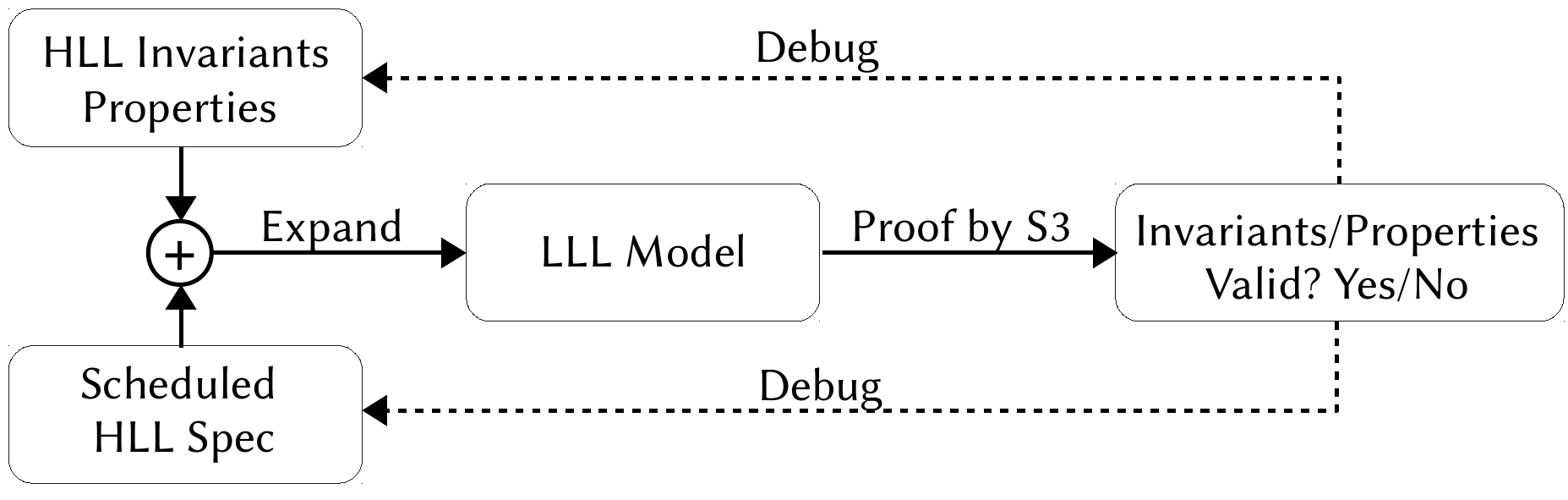}
       	\caption{Process of Property Proof using S3}
       	\label{fig:proof}
       	\end{center}
    \end{figure}

Figure \ref{fig:proof} presents the process of property proof using S3. The HLL model, combined with properties expressed in HLL as well, are expanded to a LLL model that is fed to the S3-core. If a property is falsifiable, a generated counterexample can be simulated at the HLL level to help the debug. S3 also provides automatic analysis tools to help the search of lemmas used by the proof. 

\section{Translating HLL to HLL-Equivalent C Code}
\label{sec:hll2c}
Once the Event-B model is translated into a HLL model, the correctness of which is proven, the next phase consists of translating the HLL model into C code and proving that the code is correct. Translation of HLL to C involves two activities: an automatic one where the C code is generated directly from the HLL model, and a manual one where some functions in the C code are implemented manually from contracts expressed in HLL. 
In this section, we define the architecture of generated C code in Section \ref{sec:arch_code}. The code generation/implementation rules from the HLL model to the C code are respectively defined in Section \ref{sec:gen_code} and \ref{sec:impl_code}. The approach of equivalence proof is briefly addressed in Section \ref{sec:equiv_proof}. Note that, instead of equivalence proof, the correctness of the code can also be guaranteed by proving the same properties (already expressed in HLL) in HLL. In order to re-prove the same properties, the C code is translated into an HLL model, where the HLL properties are combined and proved.
	
\subsection{Architecture of the Generated C Code}	
\label{sec:arch_code}
\begin{figure}[!ht]
	\begin{center}
	\includegraphics[width=12cm]{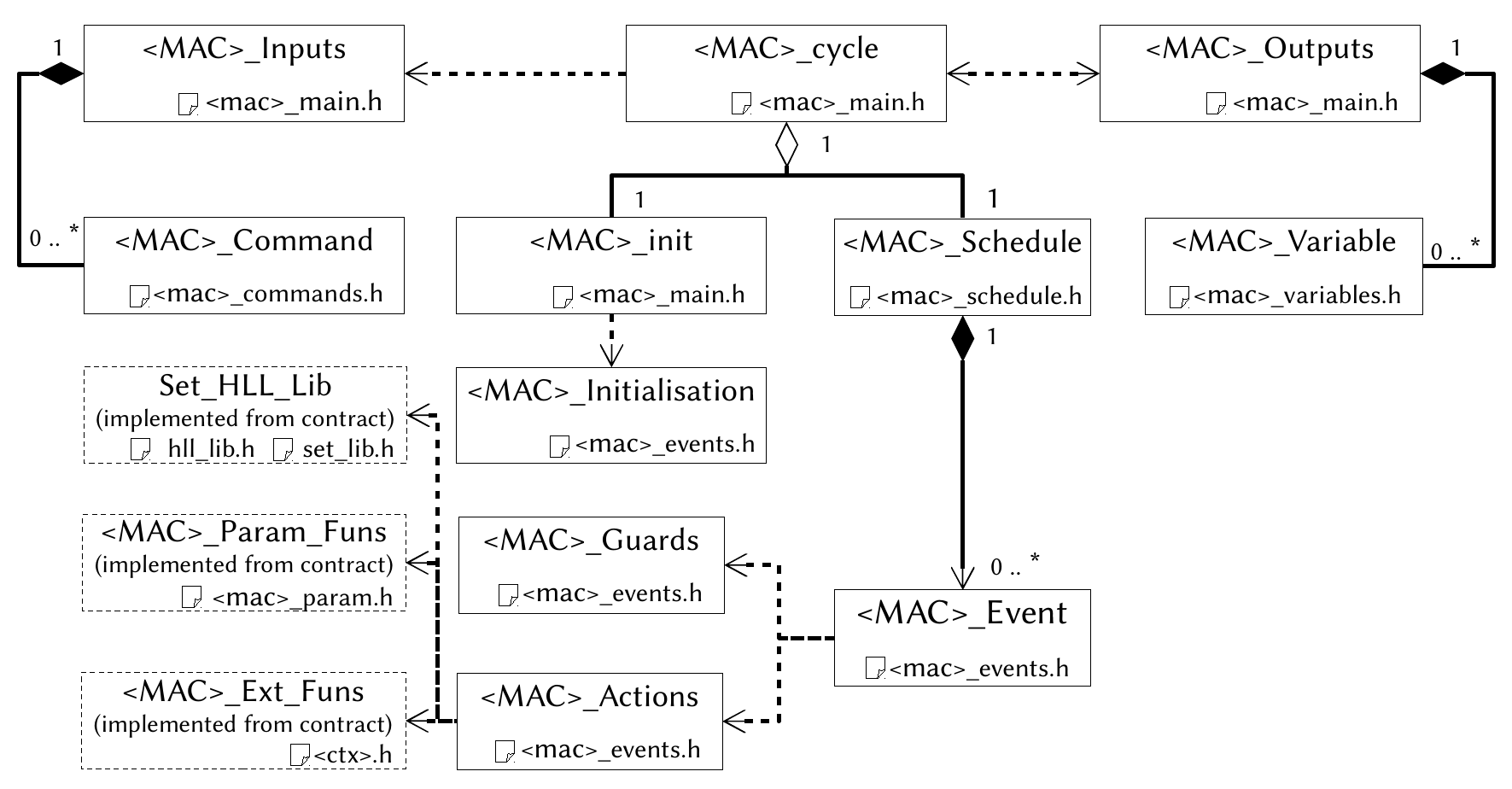}
    \caption{Static Architecture of Generated C Code}
    \label{fig:arch_code}
    \end{center}
\end{figure}

The static architecture of the C code is given on Figure \ref{fig:arch_code}. It reflects directly the architecture of the HLL model shown in Fig. \ref{fig:arch_hll} for the parts related to implementation (the white blocks).  
The dynamic architecture is described as follows: 
\begin{itemize}
\item
	The system is first initialized by the function $\mathtt{\langle MAC \rangle\_init}$ that calls the function $\mathtt{\langle MAC \rangle\_Initialisation}$, 
\item
	Then the system calls periodically the scheduler $\mathtt{\langle MAC \rangle\_Schedule}$. The scheduler calls each event processing functions $\mathtt{\langle MAC \rangle\_Events}$ according to the schedule order. Each event processing function evaluates its guards $\mathtt{\langle MAC \rangle\_Guards}$ and, if all guards are enabled and the event is triggered, call the action realization function $\mathtt{\langle MAC \rangle\_Actions}$. The guards and actions depend on the inputs $\mathtt{\langle MAC \rangle\_Inputs}$, outputs $\mathtt{\langle MAC \rangle\_Outputs}$ and the functions implemented from the contracts.
\end{itemize}

\subsection{Translation of HLL Model to C code}
\label{sec:gen_code}
The C code can be generated either from the HLL model or from the LLL model that is the expanding result of the HLL model. The alternative is presented and discussed hereafter.  

\subsubsection{Code Generation from HLL}
\begin{figure}[!ht]
	\begin{center}
	\includegraphics[width=12.5cm]{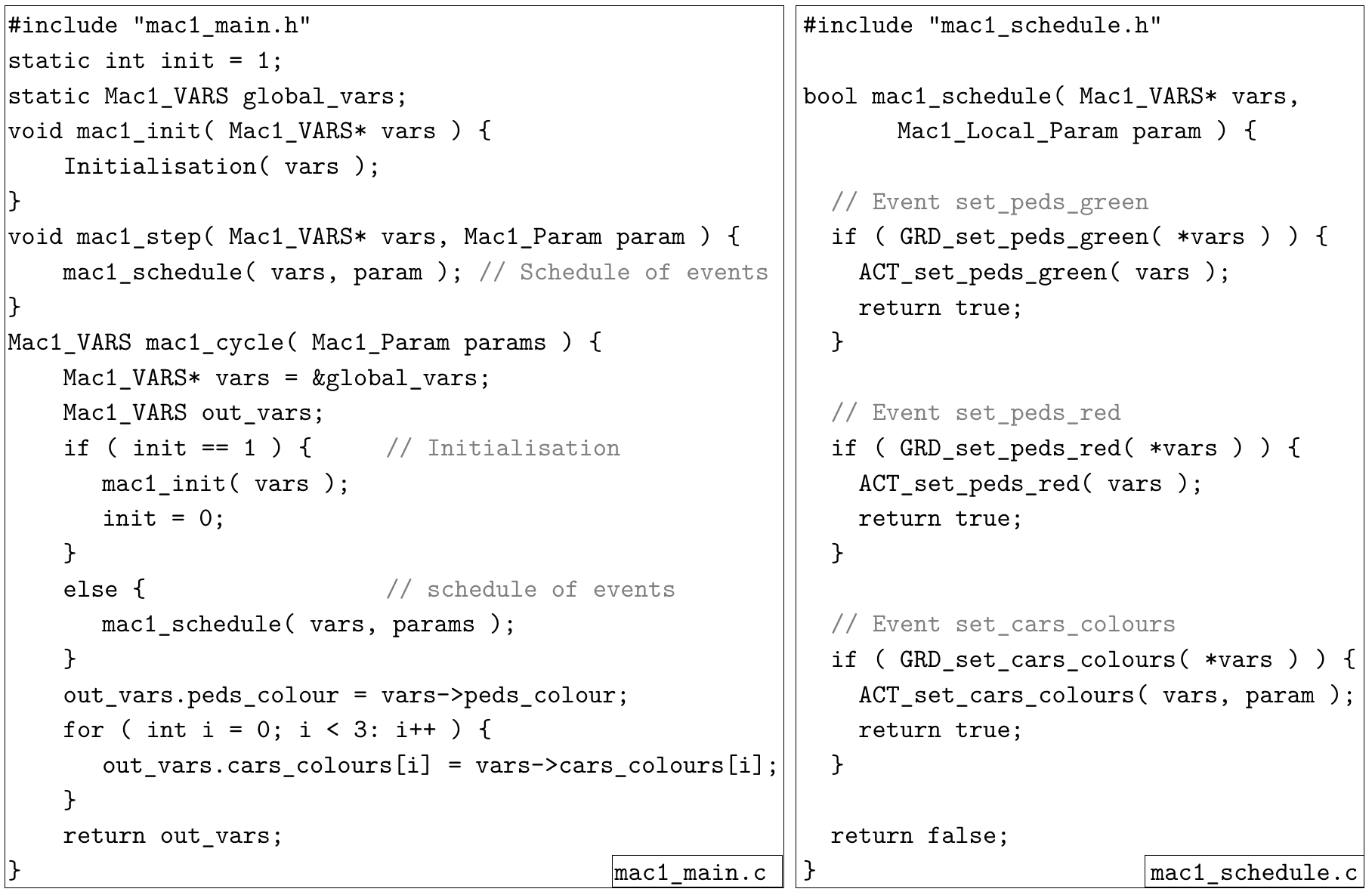}
    \caption{Generated C Code of the Running Example}
    \label{fig:mac1_c}
    \end{center}
\end{figure}

Generating C code from HLL is in a way similar to that of the translating Event-B to HLL. Figure \ref{fig:mac1_c} shows the generated C code $\mathtt{mac1\_main.c}$ and $\mathtt{mac1\_schedule.c}$ of the running example. They are generated from the HLL model $\mathtt{Mac1}$ in Fig. \ref{fig:mac1_hll}. The schedule defined in the $\mathtt{Mac1}$ HLL model is extracted in the code $\mathtt{mac1\_schedule.c}$. The main file $\mathtt{mac\_main.c}$ conforms to the cyclic execution in the HLL model. A set of global variables is defined as the set of system states and initialized. They are cyclically updated according to the order defined by the event scheduler. The guards and actions are defined as functions in the $\mathtt{mac1\_events.c}$. Because of space limitations, the complete C code of the running example is provided in the Appendix C.   

\subsubsection{Code Generation from LLL}
\begin{figure}[!ht]
	\begin{center}
	\includegraphics[width=10cm]{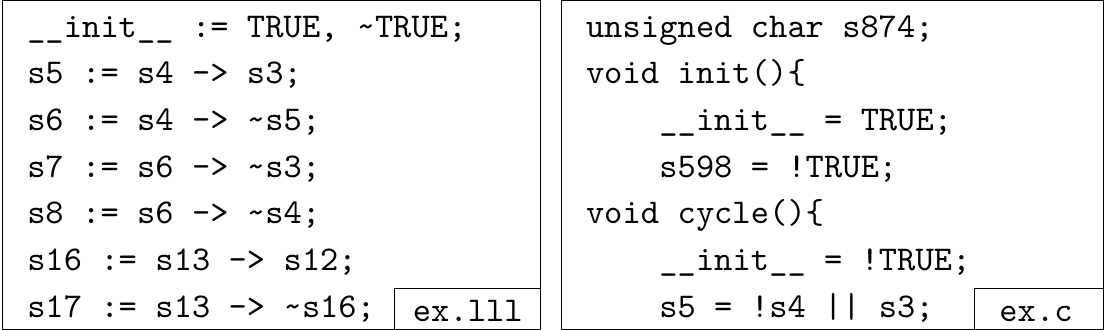}
    \caption{Generated C Code from the LLL Model}
    \label{fig:lll_c}
    \end{center}
\end{figure}

The C code may also be generated directly from the intermediate LLL model, which is the expanded translation of the HLL model. This is achieved by the LLL to C translator in S3. 
We present an example in Figure \ref{fig:lll_c}. In the model $\mathtt{ex.lll}$, all variables are boolean, and only three bitwise operators are used, therefore the translation is direct. The C code $\mathtt{ex.c}$ contains a $\mathtt{init}$ function and a $\mathtt{cycle}$ function. The translation is automatic, and the cost of the conversion from LLL to C is very low. However, as all operations - including arithmetic operations - are encoded using bit-level C operators, execution performance degrades gradually as the number of arithmetic operations increases. This solution is interesting for applications that perform mainly logical operations. 

\subsection{Code Implementation from HLL Contracts}
\label{sec:impl_code}
\begin{figure}[ht!]
	\begin{center}
	\includegraphics[width=11.5cm]{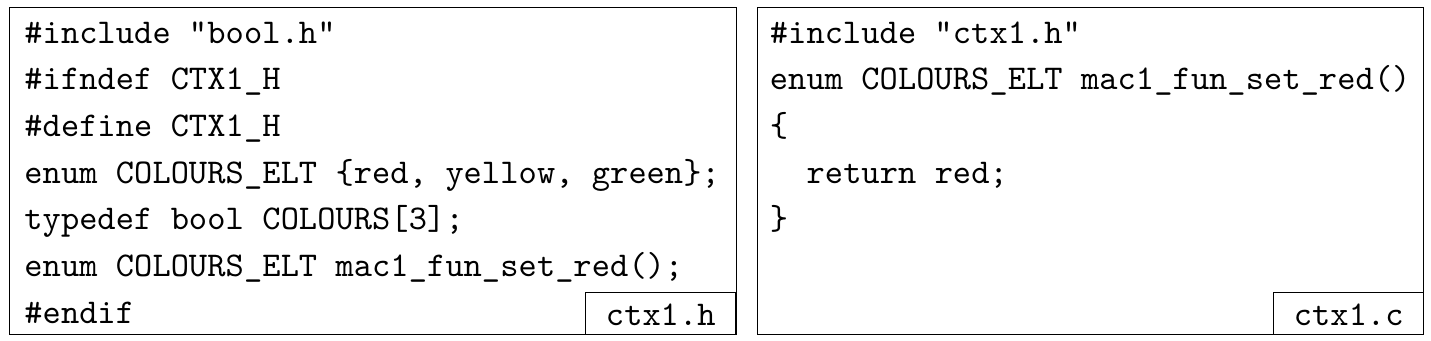}
    	\caption{Example of Implementation from HLL Contracts}
    	\label{fig:fun_contract}
    	\end{center}
\end{figure}

The HLL function contracts require a developer to translate it to C. This is the usual practice, even though in our case, the software specification is formal. We show a function implemented under the HLL contract used in the running example in Figure \ref{fig:fun_contract}. The function $\mathtt{mac1\_fun\_set\_red()}$ is implemented from the contract expressed as constraint on the block $\mathtt{fun\_set\_red()}$ in the namespace $\mathtt{ctx1}$ in Fig. \ref{fig:mac1_hll}, shown as follows.
\begin{align*}
\mathtt{ALL \enspace f: fun\_set\_red \enspace (fun\_vars.Contract\_input\_fun\_set\_red = red);}
\end{align*}

\subsection{Proving Equivalence between HLL Model and C Code}
\label{sec:equiv_proof}
\begin{figure}[ht!]
	\begin{center}
	\includegraphics[width=9cm]{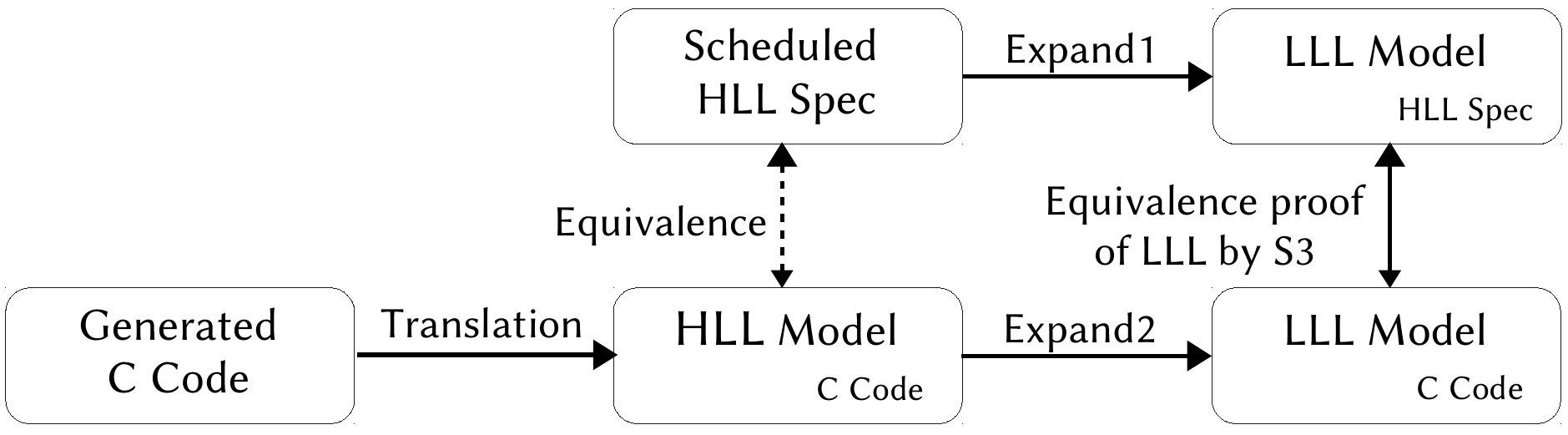}
    	\caption{Process of Equivalence Proof}
    	\label{fig:equiv_proof}
    	\end{center}
\end{figure}

Proving the equivalence of two models (or one program and one model) consists in proving that the two systems behave identically (in particular, provide the same outputs) for any input in the input domain. In our case, we rely on the equivalence proof to guarantee the correctness of the code. 
The equivalence proof is concerned with two problems: the equivalence between HLL model and the generated C code, and the equivalence between HLL contracts and the implemented C code. Figure \ref{fig:equiv_proof} presents the process of proving the equivalence between the HLL model and the generated/implemented C code. The C code is translated into another HLL model. Both HLL models are then respectively expanded to two LLL models using diversified expanders\footnote{The diversified expanders have been designed and implemented by different teams using different programming languages.}. To prove the equivalence of two HLL models, the equivalence models are constructed and proved at the LLL level.

The approach is sound for two reasons. First, the correctness of the HLL model of the system is proven,  as shown in Section \ref{sec:b2hll}. Second, the verification means (i.e., the S3 toolset) is developed so as to facilitate qualification against certification standards (in particular, DO-178C \cite{do2011178c} and DO-330 \cite{do2011330}). Towards this goal, S3 is organized in a set of small, independent components, from which the most critical ones - an equivalence model constructor, and a tool to verify the validity of the proof - are developed according to the highest integrity levels.

In practice, it shall be noted that some of the Event-B events actually model the modifications of monitored and commanded variables, i.e., variables that are not modified by the function under design \cite{parnas1991functional,butler2009towards}.
The actions triggered by these events describe the expected effects of the external environment on these variables. As they do not represent actions of the system, they shall not be translated to software code. Thus, they are implemented by simple interface functions performing acquisition of external variables. 
As a result, no equivalence checking can be done for those parts of the generated C code, their verification shall thus be achieved using other means (formal or informal). 

\section{The Case Study: Automatic Rover Protection}
\label{sec:arp}    
\begin{figure}[ht!]
	\begin{center}
	\includegraphics[width=12cm]{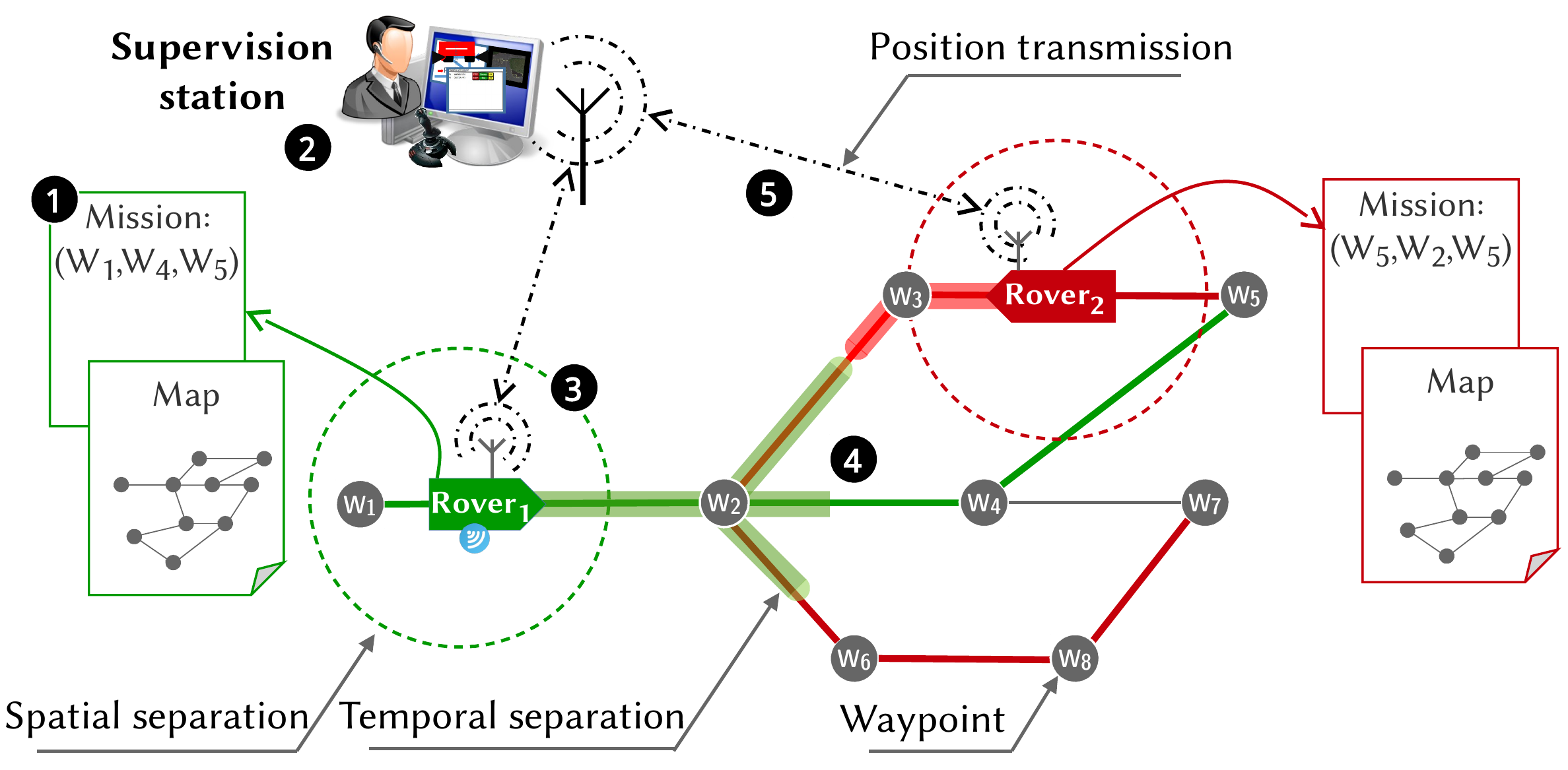}
    \caption{ARP System Overview}
    \label{fig:arp}
    \end{center}
\end{figure}

TwIRTee is the small three-wheeled robot (or “rover”) used as the demonstrator of the INGEQUIP project\footnote{The INGEQUIP project is conducted at the Institut de Recherche Technologique of Toulouse (IRT-Saint Exupéry)}. It is used to evaluate new methods and tools in the domain of hardware/software co-design, virtual integration, and application of formal methods for the development of equipments. TwIRTee’s architecture, software, and hardware components are representative of a significant family of aeronautical, spatial and automotive systems \cite{cuenot2016experiments}.
A rover performs a sequence of missions (\circled{\textbf{1}} on Figure \ref{fig:arp}). A mission is defined by a start time and an ordered set of waypoints to be passed-by. Missions are planned off-line and transmitted to the rover by a supervision station (\circled{\textbf{2}}). To go from the first waypoint to the last, the rover moves on a track that is materialized by a grey line on the ground. In a more abstract way, a complete mission can be modelled by a path in a graph where nodes represent waypoints, and edges represent parts of the track joining two waypoints. 
A rover shares the tracks with several identical rovers. In order to prevent collisions, each of them embeds a protection function (or ARP) which purpose is to maintain some specified spatial (\circled{\textbf{3}}) and temporal separation (\circled{\textbf{4}}) between them. On Figure 1, temporal separations are represented by light green and light red areas superimposed on the map: basically, rover$_{2}$ (resp. rover$_1$) shall never enter the light green (resp. light red).
In the current implementation, the ARP essentially acts by reducing the rover speed and, in some specific cases, by performing a simple avoidance trajectory.  To take the appropriate action, the ARP exploit the following pieces of information: the map, the position of all other rovers transmitted by a centralized supervision station (\circled{\textbf{5}}), and its own position. 

For this paper, we rely on a specific model of the ARP function where some elements have been simplified. We thus only consider specification elements such as rover position, speed, deceleration, and others as being discrete values (no use of Real or Floating Point data). The statistics of the Event-B model, translated HLL models, and generated/implemented C code are provided in Table \ref{table:arp_statistics}. 
\begin{table}[!ht]
\caption{Statistics of the ARP Models (Event-B and HLL) and Code}
\centering
\footnotesize
\begin{tabular}{ l |  l | c }
\multirow{4}{*}{Event-B model} & Events / Actions / Guards & 18 / 70 / 74 \\
& Axioms / Theorems / Constants / Variables / Invariants & 40 / 11 / 27 / 15 / 79   \\
& PO (Total / Automatic / Manuel) & 634 / 626 / 8 \\
& LOC of Event-B model of the final refinement &  243 \\
\hline
\multirow{4}{*}{HLL models} & Properties ( safety / liveness/ deadlock-freeness ) & 42 / 4 / 1 \\
& Time for property proof  ( invariants + properties ) & 4s \\
& Contracts ( external function / parameter ) & 5 / 12 \\
& LOC of HLL models ( system / verification ) & 800 / 500 \\
\hline
\multirow{2}{*}{C code} & LOC of C code(generated / implemented )& 1200 / 600 \\
& Time for equivalence proof & 3s \\
\end{tabular}
\label{table:arp_statistics}
\end{table}

\section{Conclusion and Perspective}
\label{sec:conclusion}
\subsection{Conclusion}
This work addresses the translation of Event-B into C and the demonstration of the correctness of translation using formal methods. Our approach relies on an intermediate modeling/verification language HLL. The correctness proof follows two steps. First, the final refinement of the Event-B model, including invariants from all refined machines and elements from all extended contexts, are translated to the HLL model. Additional properties are expressed in HLL. The invariants/properties are proved at the HLL level to guarantee the correctness of the translation. Second, the C code is automatically generated from the HLL model for most of the system functions and manually for implemented from the contracts for the remaining ones. The equivalence between the code and the HLL model is proved to guarantee the correctness of the code. In this paper, we define the translation rules, and show the experimented results on a significant use case. More details about the refinement and design can be found in \cite{arnaud2017}. 

Compared to the existing approaches to proving the direct translation from the Event-B model to generated code, our contribution is to address the verification of additional properties (i.e., deadlock-freeness and liveness properties) in the HLL model. For the direct translation approaches \cite{mery2011automatic,meryM13transformation,furst2014code}, the analysis of these properties needs to be performed in the final C code. Considering the complexity of analysis, the authors of these works all mentioned the limits of their approaches and considered it as an open issue. 

Another advantage of our approach is the support of property proof and equivalence proof of Floating-Point Arithmetic (FPA) in both the HLL model and the code. The S3 toolset provides an HLL library of FPA based on bit-blasting\footnote{Bit-blasting is a classic method that translates bit-vector formulas into propositional logic expressions.} \cite{clabaut2016industrial} that conforms to the FPA standard \textit{IEEE std 2008-754} \cite{zuras2008ieee}.

\subsection{Perspectives}
The development of the translation tool is ongoing. As the specification of the HLL language will be published in a near future, it will soon become possible to integrate the translator as a plug-in in the Rodin platform.

The antecedent of Event-B, the B method, supports formal refinement until the programming language B0, that is translated later to C/ADA code. This method requires more refinements, but the resulting B0 model is sufficiently close to software code that code generation becomes straightforward. The correctness of final code still needs to be proved. As the verification gap between B0 and C code is smaller than the one between Event-B and C code, the results of this work  could be adapted to the translation from B0 to C via HLL. 

\section*{Acknowledgments}
\label{sec:ack}
This work has been funded by the INGEQUIP project. The authors would like to thank the members in the project, and the colleagues of the Systerel S.A.S company Nicolas Breton, Mathieu Clabaut and Yoann Fonteneau for their good cooperation. 
The author Ning Ge would like to thank Hongyu Liu for his help on this work. 

	\bibliographystyle{plain}
\bibliography{tacas2017}



\section*{Supplementary material}

# Appendix A: Event-B Model of Traffic Light Control

**CONTEXT** ctx1
**SETS**
    COLOURS
**CONSTANTS**
    red
    yellow
    green
    fun_set_red
**AXIOMS**
    axm1: $partition(COLOURS, \{red\}, \{yellow\}, \{green\})$
    axm2: $fun\_set\_red \in COLOURS$
    axm3: $fun\_set\_red = red$
**END**

---

**MACHINE** mac0
**VARIABLES**
    cars_go
    peds_go
**INVARIANTS**
    inv1: $cars\_go \in BOOL$
    inv2: $peds\_go \in BOOL$
    inv3: $\neg(cars\_go = TRUE \land peds\_go = TRUE)$
**EVENTS**
**Initialisation**
    **begin**
        act1: $cars\_go := FALSE$
        act2: $peds\_go := FALSE$
    **end**
**Event** set_peds_go ⟨ordinary⟩ $\widehat{=}$
    **when**
        grd1: $cars\_go = FALSE$
    **then**
        act1: $peds\_go := TRUE$
    **end**
**Event** set_peds_stop ⟨ordinary⟩ $\widehat{=}$
    **begin**
        act1: $peds\_go := FALSE$
    **end**
**Event** set_cars ⟨ordinary⟩ $\widehat{=}$
    **any**
        new_value
    **where**
        grd1: $new\_value \in BOOL$
        grd2: $new\_value = TRUE \Rightarrow peds\_go = FALSE$
    **then**
        act1: $cars\_go := new\_value$
    **end**
**END**

**MACHINE** mac1
**REFINES** mac0
**SEES** ctx1
**VARIABLES**
    peds_colour
    cars_colours
**INVARIANTS**
    inv4: $peds\_colour \in \{red, green\}$
    inv5: $cars\_colours \subseteq COLOURS$
    gluing_peds: $peds\_go = TRUE \Leftrightarrow peds\_colour = green$
    gluing_cars: $cars\_go = TRUE \Leftrightarrow green \in cars\_colours$
**EVENTS**
**Initialisation**
    **begin**
        act1: $peds\_colour := red$
        act2: $cars\_colours := \{red\}$
    **end**
**Event** set_peds_green ⟨ordinary⟩ ≙
**refines** set_peds_go
    **when**
        grd1: $green \notin cars\_colours$
    **then**
        act1: $peds\_colour := green$
    **end**
**Event** set_peds_red ⟨ordinary⟩ ≙
**refines** set_peds_stop
    **begin**
        act1: $peds\_colour := fun\_set\_red$
    **end**
**Event** set_cars_colours ⟨ordinary⟩ ≙
**refines** set_cars
    **any**
        new_value_colours
    **where**
        grd1: $new\_value\_colours \subseteq COLOURS$
        grd2: $green \in new\_value\_colours \Rightarrow peds\_colour = red$
        grd3: $cars\_colours = \{yellow\} \Rightarrow new\_value\_colours = \{red\}$
        grd4: $cars\_colours = \{red\} \Rightarrow new\_value\_colours = \{red, yellow\}$
        grd5: $cars\_colours = \{red, yellow\} \Rightarrow new\_value\_colours = \{green\}$
        grd6: $cars\_colours = \{green\} \Rightarrow new\_value\_colours = \{yellow\}$
    **with**
        new_value: $new\_value = TRUE \Leftrightarrow green \in new\_value\_colours$
    **then**
        act1: $cars\_colours := new\_value\_colours$
    **end**
**END**

# Appendix B: HLL Models of Traffic Light Control

```
Namespaces:
Ctx1 {
Types:
    enum {red, yellow, green} COLOURS_ELT;
    bool COLOURS[3];   /* Ordered array: 0-red, 1-yellow, 2-green */
Types:
    struct {         /* Variables for specifying function contracts */
        Contract_input_fun_set_red : COLOURS_ELT
    } Fun_Contract_Vars;
Declarations:        /* Variables of function contract */
    Fun_Contract_Vars fun_vars;
Blocks:              /* mac1/set_peds_red/fun_set_red() */
fun_set_red() -> (COLOURS_ELT colour) {
    colour := fun_vars.Contract_input_fun_set_red;
}
Constraints:         /* Contract of function set_peds_red() */
    ALL f: fun_set_red (f.colour = red);
}

_____

Mac0 {
Types:
    struct {   /* PARAMETERS */
        set_cars_new_value : bool
    } Param;
    struct {   /* Gluing VARIABLES in mac0 */
        cars_go : bool,
        peds_go : bool
    } VARS;
Inputs:     /* VARIABLES in mac0 */
    VARS vars;
    Param param;
}

_____

Mac0_INV {
Proof Obligations:
    /* INVARIANTS in mac0: mac0/inv3  */
    ~(Mac0::vars.peds_go & Mac0::vars.cars_go);
}

_____

Mac1_Mac0_Gluing_INV {
Constraints:
    /* Gluing invariants */
    Mac0::vars.peds_go <-> Mac1::peds_colour = Ctx1::green;
    Mac0::vars.cars_go <-> Mac1::cars_colours[2];

    /* Witness in mac1/set_cars_colours */
    Mac0::param.set_cars_new_value <-> Mac1::param.set_cars_colours_new_value_colours[2];
}

_____

Mac1 {
Types:
    struct {         /* VARIABLES in mac1 */
        peds_colour : Ctx1::COLOURS_ELT,
        cars_colours: Ctx1::COLOURS
    } VARS;
    struct {         /* PARAMETERS */
        set_cars_colours_new_value_colours: Ctx1::COLOURS
    } Param;
Inputs:
    Param param;     /* Parameters */
Declarations:
    /* VARIABLES in mac1 */
    VARS vars;
    Ctx1::COLOURS_ELT peds_colour;
    Ctx1::COLOURS     cars_colours;
```

```
  /* GUARDS in mac1 */
  bool GRD_set_peds_green;
  bool GRD_set_peds_red;
  bool GRD_set_cars_colours;
Definitions:
  /* Define guards of events */
  GRD_set_peds_green  := (~vars.cars_colours[2]);
  GRD_set_peds_red    := TRUE;
  GRD_set_cars_colours:= TRUE;

  /* Define actions of events */
  /* Initialisation */
  I(peds_colour) := Ctx1::red;
  I(cars_colours):= {TRUE, FALSE, FALSE};

  /* Schedule */
  X(peds_colour) := if   GRD_set_peds_green   then Ctx1::green
                    elif GRD_set_peds_red     then Ctx1::fun_set_red()
                    elif GRD_set_cars_colours then peds_colour
                    else                           peds_colour;
  X(cars_colours):= if   GRD_set_peds_green   then cars_colours
                    elif GRD_set_peds_red     then cars_colours
                    elif GRD_set_cars_colours then param.set_cars_colours_new_value_colours
                    else                           cars_colours;
  vars := {peds_colour, cars_colours};
Outputs: vars;
}

_____

Mac1_INV {
Proof Obligations:
 /* INVARIANTS in mac1 : mac1/inv1      */
 Mac1::vars.peds_colour = Ctx1::red # Mac1::vars.peds_colour = Ctx1::green;
}

Mac1_Para_Contract {
Constraints: /* Local parameter contracts */
 /* Event set_cars_colours: grd2 */
 Mac1::param.set_cars_colours_new_value_colours[2] -> Mac1::peds_colour = Ctx1::red;
 /* Event set_cars_colours: grd3 */
 (~Mac1::cars_colours[0] & Mac1::cars_colours[1] & ~Mac1::cars_colours[2])
    -> (    Mac1::param.set_cars_colours_new_value_colours[0]
         & ~Mac1::param.set_cars_colours_new_value_colours[1]
         & ~Mac1::param.set_cars_colours_new_value_colours[2]);
 /* Event set_cars_colours: grd4 */
 (Mac1::cars_colours[0] & ~Mac1::cars_colours[1] & ~Mac1::cars_colours[2])
    -> (  Mac1::param.set_cars_colours_new_value_colours[0]
         & Mac1::param.set_cars_colours_new_value_colours[1]
         & ~Mac1::param.set_cars_colours_new_value_colours[2]);
 /* Event set_cars_colours: grd5 */
 (Mac1::cars_colours[0] & Mac1::cars_colours[1] & ~Mac1::cars_colours[2])
    -> (  ~Mac1::param.set_cars_colours_new_value_colours[0]
         & ~Mac1::param.set_cars_colours_new_value_colours[1]
         & Mac1::param.set_cars_colours_new_value_colours[2]);
 /* Event set_cars_colours: grd6 */
 (~Mac1::cars_colours[0] & ~Mac1::cars_colours[1] & Mac1::cars_colours[2])
    -> ( ~Mac1::param.set_cars_colours_new_value_colours[0]
         & ~Mac1::param.set_cars_colours_new_value_colours[1]
         & ~Mac1::param.set_cars_colours_new_value_colours[2]);
}

_____

Mac1_PROPS {
Declarations:       /* Property of deadlock freeness */
  bool PROP_DLF;
Definitions:        /* Compute property of deadlock freeness */
  PROP_DLF := TRUE,   Mac1::GRD_set_peds_green
                    # Mac1::GRD_set_peds_red
                    # Mac1::GRD_set_cars_colours;
Proof Obligations:  /* Property of deadlock freeness */
  PROP_DLF;
}
```

# Appendix C: C Code of Traffic Light Control

**ctx1.h**
```c
#include "bool.h"

#ifndef CTX1_H
#define CTX1_H
enum COLOURS_ELT {
    red,
    yellow,
    green
};
typedef bool COLOURS[3];
enum COLOURS_ELT mac1_fun_set_red();
#endif
```
_____

**ctx1.c**
```c
#include "ctx1.h"
enum COLOURS_ELT mac1_fun_set_red()
{
        return red;
}
```
_____

**mac1_events.h**
```c
#include "ctx1.h"
#include "mac1_param.h"
#include "mac1_variables.h"

#ifndef MAC1_EVENTS_H
#define MAC1_EVENTS_H

/* Guards in mac1 */
bool GRD_set_peds_green( Mac1_VARS );
bool GRD_set_peds_red( Mac1_VARS );
bool GRD_set_cars_colours( Mac1_VARS );

/* Actions in mac1 */
void Initialisation( Mac1_VARS* );
void ACT_set_peds_green( Mac1_VARS* );
void ACT_set_peds_red( Mac1_VARS* );
void ACT_set_cars_colours( Mac1_VARS* ,Mac1_Param );
#endif
```
_____

**mac1_events.c**
```c
#include "mac1_events.h"
/* Guards in mac1 */
bool GRD_set_peds_green( Mac1_VARS vars ) {
    return ( vars.cars_colours[2] == false );
}

bool GRD_set_peds_red( Mac1_VARS vars ) {
    return true;
}

bool GRD_set_cars_colours( Mac1_VARS vars ) {
    return true;
}

/* Actions in mac1 */
void Initialisation( Mac1_VARS* vars ) {
    vars->peds_colour = red;
    vars->cars_colours[0] = true;
    vars->cars_colours[1] = false;
    vars->cars_colours[2] = false;
}

void ACT_set_peds_green( Mac1_VARS* vars ) {
    Mac1_VARS pre_vars = *vars;
    vars->peds_colour =  green;
```

```c
    vars->cars_colours[0] = pre_vars.cars_colours[0];
    vars->cars_colours[1] = pre_vars.cars_colours[1];
    vars->cars_colours[2] = pre_vars.cars_colours[2];
}

void ACT_set_peds_red( Mac1_VARS* vars ) {
    Mac1_VARS pre_vars = *vars;
    vars->peds_colour = mac1_fun_set_red();
    vars->cars_colours[0] = pre_vars.cars_colours[0];
    vars->cars_colours[1] = pre_vars.cars_colours[1];
    vars->cars_colours[2] = pre_vars.cars_colours[2];
}

void ACT_set_cars_colours( Mac1_VARS* vars, Mac1_Param param ) {
    Mac1_VARS pre_vars = *vars;
    vars->peds_colour = pre_vars.peds_colour;
    vars->cars_colours[0] = param.set_cars_colours_new_value_colours[0];
    vars->cars_colours[1] = param.set_cars_colours_new_value_colours[1];
    vars->cars_colours[2] = param.set_cars_colours_new_value_colours[2];
}
```

---

**mac1_param.h**

```c
#include "ctx1.h"

#ifndef MAC1_PARAM_H
#define MAC1_PARAM_H
typedef struct {
    COLOURS set_cars_colours_new_value_colours;
} Mac1_Param;
#endif
```

---

**mac1_variables.h**

```c
#include "ctx1.h"

#ifndef MAC1_VARIABLES_H
#define MAC1_VARIABLES_H
/* VARIABLES in mac1 */
typedef struct {
    enum COLOURS_ELT peds_colour;
    COLOURS cars_colours;
} Mac1_VARS;
#endif
```

---

**mac1_schedule.h**

```c
#include "mac1_events.h"
#include "mac1_param.h"
#include "mac1_variables.h"

#ifndef MAC1_SCHEDULE_H
#define MAC1_SCHEDULE_H
bool mac1_schedule( Mac1_VARS*, Mac1_Param );
#endif
```

---

**mac1_schedule.c**

```c
#include "mac1_schedule.h"
bool mac1_schedule( Mac1_VARS* vars, Mac1_Param param ) {
        // Event set_peds_green
        if ( GRD_set_peds_green( *vars ) )
        {
                ACT_set_peds_green( vars );
                return true;
        }

        // Event set_peds_red
        if ( GRD_set_peds_red( *vars ) )
        {
                ACT_set_peds_red( vars );
                return true;
```

```c
        }

        // Event set_cars_colours
        if ( GRD_set_cars_colours( *vars ) )
        {
                ACT_set_cars_colours( vars, param );
                return true;
        }

        return false;
}
```

---

**mac1_main.h**

```c
#include "ctx1.h"
#include "mac1_param.h"
#include "mac1_variables.h"
#include "mac1_events.h"
#include "mac1_schedule.h"

#ifndef MAC1_MAIN_H
#define MAC1_MAIN_H
typedef struct {
    Mac1_Param mac1_param;
} Mac1_Inputs;

typedef struct {
    Mac1_VARS mac1_vars;
} Mac1_Outputs;

void mac1_init( Mac1_VARS* );
void mac1_step( Mac1_VARS*, Mac1_Param );
Mac1_VARS mac1_cycle( Mac1_Param );
#endif
```

---

**mac1_main.c**

```c
#include "mac1_main.h"
static int init = 1;
static Mac1_VARS global_vars;

void mac1_init( Mac1_VARS* vars ) {
    Initialisation( vars );
}

void mac1_step( Mac1_VARS* vars, Mac1_Param param ) {
    mac1_schedule( vars, param );
}

Mac1_VARS mac1_cycle( Mac1_Param params ) {
    Mac1_VARS* vars = &global_vars;
    Mac1_VARS out_vars;

    if ( init == 1 ) {
            mac1_init( vars );
            init = 0;
    }
    else {
            mac1_step( vars, params );
        }

    out_vars.peds_colour = vars->peds_colour;
    for ( int i = 0; i < 3: i++ ) {
            out_vars.cars_colours[i] = vars->cars_colours[i];
    }

    return out_vars;
}
```


\label{appendix:c}

\end{document}